\documentclass[floatfix,aps,prd,twocolumn,superscriptaddress,groupedaddress,nofootinbib
]{revtex4} 

\usepackage{amsfonts,amsmath,amssymb}
\usepackage{graphicx}
\usepackage{xspace}
\usepackage[caption=false]{subfig}
\usepackage{hyperref}
\usepackage[utf8]{inputenc}
\newcommand{\z}[1]{\ensuremath{\mathbb{Z}_#1}}

\newcommand{\gev}{\ensuremath{\,\text{GeV}}}

\newcommand{\MP}{\ensuremath{M_P}\xspace}
\newcommand{\MI}{\ensuremath{\Lambda_\text{I}}\xspace}

\newcommand{\order}[1]{\ensuremath{\mathcal{O}\left(#1\right)}}

\newcommand{\vev}[1]{\ensuremath{\langle#1\rangle}}

\newcommand{\code}{\texttt}

\newcounter{comment}

\newcommand{\deriv}[2]{\frac{\text{d}#1}{\text{d}#2}}
\newcommand{\pderiv}[2]{\frac{\partial #1}{\partial #2}}
\newcommand{\dint}{\text{d}}

\newcommand{\eg}{e.g.,\xspace}
\newcommand{\ie}{i.e.,\xspace}

\newcommand{\se}{\ensuremath{\mathcal{S}_\text{E}}}

\let\oldcite\cite
\renewcommand{\cite}{\unskip~\oldcite}
\newcommand{\reftable}[1]{Table~\ref{#1}}
\newcommand{\reffig}[1]{Fig.~\ref{#1}}
\newcommand{\refeq}[1]{Eq.~(\ref{#1})}
\newcommand{\refsec}[1]{section~\ref{#1}}
\newcommand{\refcite}[1]{Ref.\cite{#1}}
\newcommand{\refapp}[1]{Appendix~\ref{#1}}
\newcommand{\see}[1]{(see \eg\refcite{#1})}

\begin{document}

\title{Gravitational waves at aLIGO and vacuum stability with a scalar singlet extension of the Standard Model}

\author{Csaba Bal\'azs}
\email{csaba.balazs@monash.edu}
\affiliation{ARC Centre of Excellence for Particle Physics at the Tera-scale, School of Physics and Astronomy, Monash University, Melbourne, Victoria 3800 Australia}
\affiliation{Monash Centre for Astrophysics, School of Physics and Astronomy, Monash University, Melbourne, Victoria 3800 Australia}

\author{Andrew Fowlie}
\email{andrew.fowlie@monash.edu}
\affiliation{ARC Centre of Excellence for Particle Physics at the Tera-scale, School of Physics and Astronomy, Monash University, Melbourne, Victoria 3800 Australia}

\author{Anupam Mazumdar}
\email{a.mazumdar@lancaster.ac.uk}
\affiliation{Consortium for Fundamental Physics, Physics Department, Lancaster University,
LA1 4YB, UK}
\affiliation{Kapteyn Astronomical Institute, University of Groningen, 9700 AV Groningen, The Netherlands}

\author{Graham A.~White}
\email{graham.white@monash.edu}
\affiliation{ARC Centre of Excellence for Particle Physics at the Tera-scale, School of Physics and Astronomy, Monash University, Melbourne, Victoria 3800 Australia}

\date{\today}

\begin{abstract}
A new gauge singlet scalar field can undergo a strongly first-order phase transition (PT) leading to gravitational waves (GW) potentially observable at aLIGO and simultaneously stabilize the electroweak vacuum. 
aLIGO (O5) is potentially sensitive to cosmological PTs at $10^7$-$10^8\gev$, which coincides with the requirement that the singlet scale is less than the Standard Model (SM) vacuum instability scale, which is between $10^8\gev$ and $10^{14}\gev$. 
After sampling its parameter space, we identify three benchmark points with a PT at about $T\approx 10^7\gev$ in a gauge singlet extension of the SM.
We calculate the nucleation temperature, order parameter, characteristic timescale, and peak amplitude and frequency of GW from bubble collisions during the PT for the benchmarks and find that, in an optimistic scenario, GW from such a PT may be in reach of aLIGO (O5).
We confirm that the singlet stabilizes the electroweak vacuum whilst remaining consistent with zero-temperature phenomenology as well.
Thus, this scenario presents an intriguing possibility that aLIGO may detect traces of fundamental physics motivated by vacuum stability at an energy scale that is well above the reach of any other experiment.
\end{abstract}

\maketitle

\section{Introduction}

The recent detection of gravitational waves (GW) by the LIGO Collaboration opened a new observational window for the early Universe\cite{Abbott:2016blz}.  Among the most exciting prospects is the observation of GW from cosmological events that happened well before the first observable photons were created\cite{Lasky:2015lej}.  Not limited by recombination, GW can be used to directly probe fundamental physics, reaching to considerably higher energies than any other existing experiments. There are potentially several known sources of observable GW, which can be split into three categories\cite{Maggiore:1999vm}: (i)  binary black hole mergers, mergers of binary neutron stars or a neutron star and a black hole, or supernova core collapse, with a duration between a millisecond and several hours; (ii) long duration signals, \ie from spinning neutron stars; and (iii) stochastic background arising from the superposition of unresolved astrophysical sources.  The latter can be a stochastic background of GW which can also arise from cosmological events, such as during primordial inflation\cite{Grishchuk:1974ny,Starobinsky:1979ty,Rubakov:1982df}, after inflation during resonant preheating\cite{Khlebnikov:1997di,Easther:2006gt,Dufaux:2007pt,GarciaBellido:2007af,Mazumdar:2008up}, or due to fragmentation of the inflaton or any scalar condensate\cite{Kusenko:2008zm,Kusenko:2009cv,Mazumdar:2010pn}, cosmic strings\cite{Damour:2000wa,Olmez:2010bi}, and cosmological phase transitions (PT) accompanying either the breakdown of a fundamental symmetry or a scalar field acquiring a vacuum expectation value (VEV). If this PT is first order, then GW are created by violent collisions between expanding bubble walls
% SW: and sound waves from turbulence of expanding bubble walls 
of the new vacuum \see{Kosowsky:1992rz,Kosowsky:1991ua,Kamionkowski:1993fg,Maggiore:1999vm, Cutler:2002me, Grojean:2006bp, Huber:2008hg,Caprini:2007xq,arXiv:0711.2593, arXiv:0802.2452, arXiv:0911.0687, arXiv:1003.2462, arXiv:1004.2504, arXiv:1004.4187, arXiv:1011.5881, arXiv:1103.2159, arXiv:1104.5487, arXiv:1105.5283, arXiv:1201.6589, arXiv:1304.2433, arXiv:1305.3392, Vlcek:2013aha, arXiv:1311.0921, arXiv:1402.1345, arXiv:1407.2882, arXiv:1412.5147,Schwaller:2015tja, Chala:2016ykx,Jinno:2016knw,Kakizaki:2015wua,Hashino:2016rvx}, which can be potentially constrained by the current and future GW observatories, such as the future space mission eLISA\cite{Klein:2015hvg,Caprini:2015zlo}, and also possibly by aLIGO within the next 5 years\cite{Dev:2016feu}. Recently, it has been shown that these GW are detectable by  BBO or DECIGO\cite{jaeckel:2016jlh,Artymowski:2016tme, Addazi:2016fbj,Huang:2016cjm,Hashino:2016xoj}.

In the present work, we explore the detectability of GW originating from fundamental physics at the upgraded LIGO detector, aLIGO, in the near future (2020-22)\cite{Harry:2010zz, TheLIGOScientific:2014jea,Martynov:2016fzi,TheLIGOScientific:2016wyq}. It is known  that the frequency of GW from the electroweak PT is too low to be detected at aLIGO\cite{Grojean:2006bp, Huber:2008hg}. Therefore, our main emphasis here is to seek GW accompanying an earlier PT with physics beyond the Standard Model (BSM).  In search of detectable primordial GW at aLIGO (LIGO run phase O5), we provide a simple but concrete particle physics model which can yield the observed amplitude and peak frequency for GW which have been recently proposed in \refcite{Dev:2016feu}. In the current paper we analyse a framework which is an extension of the Standard Model (SM) of elementary particles with a gauge singlet scalar (SSM) \see{Ghosh:2015apa,O'Connell:2006wi}.  Indeed, this is the simplest example of BSM physics that could enhance electroweak vacuum stability\cite{Gonderinger:2009jp, EliasMiro:2012ay,Lebedev:2012zw}. Besides this, such a simple choice for physics beyond the SM could also help us understand primordial inflation\cite{Giudice:2010ka,Salvio:2015kka,Salvio:2015jgu} (for a review see \refcite{Mazumdar:2010sa}).

%In this paper we will provide a complementary test for the SSM. 
As noticed before in \refcite{Dev:2016feu}, aLIGO is potentially sensitive to cosmological PTs occurring at scales $10^7\gev$ to $10^8\gev$, which raises the question whether such a new scale emerges in BSM physics. It is a well established result that the observed values of the top-quark mass, Higgs mass and strong coupling drive the SM Higgs quartic coupling, via renormalization evolution, to negative values at about $\MI \sim 10^{10}\gev$.  The latter is known as the Higgs or vacuum instability scale\cite{EliasMiro:2011aa,Alekhin:2012py,Degrassi:2012ry,Masina:2012tz,Xing:2011aa}. The SM scalar potential is believed to be metastable; although we live in a false vacuum, the probability of tunneling to the true vacuum is negligible, and for a heavy Higgs boson, $m_h \gtrsim 130\gev$, the Higgs potential would be stable\cite{Degrassi:2012ry}. 
 
In this paper we show two important results, which we can summarize below:

\begin{itemize}

\item{It is possible to realise a successful strong first-order PT in the singlet direction with the nucleation temperature within the range of $10^{7}-10^{8}$~GeV,
which would give rise to a GW signal within the frequency range of aLIGO, \ie $10-100$~Hz. We will establish this by taking into account finite-temperature corrections, first incorporated in \refcite{Balazs:2013cia} in the context of the next-to minimal supersymmetric SM.}

\item{We carefully compute the running of the couplings in the SSM at two-loop, and conclude that for the range of parameters we have scanned, parameters that yield a strong first-order PT could also ameliorate the SM Higgs metastability.  In this paper we shall provide three benchmark points, where the scale of BSM physics would leave an undeniable footprint in the GW signal, potentially within the range of aLIGO (O5).}

\end{itemize}

Our paper is organised as follows: in \refsec{Sec:SSM}, we first explain the SSM model. In \refsec{Sec:PT}, we discuss what range of parameters the singlet can yield strong first-order PT, and what are the conditions to be fulfilled. In section \refsec{Sec:GW}, we briefly discuss GW amplitude and frequency from the first-order PT.
In \refsec{Sec:Vacuum_Stabiliity}, we discuss the Higgs vacuum stability in the presence of a singlet-Higgs interaction, and in \refsec{Sec:Numerics} we discuss our numerical results. In \refsec{Sec:Summary}, we conclude our results and discuss briefly future directions.

\section{Singlet extension of the Standard Model}\label{Sec:SSM}

We consider the SM plus a real scalar \see{Ghosh:2015apa,O'Connell:2006wi} that is a singlet under the SM gauge groups and carries no
\eg discrete charges. Thus, our model is described by the tree-level scalar potential
\begin{equation}\label{Eq:VTree}
\begin{split}
V_0(H, S) ={}& \mu^2 |H|^2 + \frac12 \lambda |H|^4\\
&+ \frac12 M_S^2 S^2 + \frac13 \kappa S^3 + \frac12 \lambda_S S^4\\
&+ \kappa_1 S |H|^2 + \frac12 \kappa_2 S^2 |H|^2,
\end{split}
\end{equation}
where $M_S$ is the mass parameter of the singlet, $\kappa$ is a dimensionful coupling, $\lambda_S$ is the singlet quartic coupling,
and $\kappa_{1,2}$ are singlet-Higgs couplings.
The above potential is the most general gauge invariant, renormalizable scalar potential with the considered particle content. 
The linear operator $m^3 S$ is removed by a shift in the singlet field without loss of generality. 

To account for changing field properties during cosmological PTs, we consider a one-loop effective potential with finite-temperature corrections (\ie a free energy). As the Universe cools the free energy develops a deeper minimum in the singlet direction, 
there is a PT to a new ground state and the singlet acquires a VEV,
although no symmetries are broken. 
If there is a discontinuity in the order parameter 
\begin{equation*}
 \gamma \equiv \vev{S} / T,   
\end{equation*}
\ie the PT is first order, bubbles spontaneously emerge in the Universe in which the singlet VEV is non-vanishing $\vev{S} \neq 0$. We will scan over the Lagrangian parameters at the high scale, guarantee that a strongly first-order PT occurs at a critical temperature in the range $(10^7, 10^8) \gev$ by solving for Lagrangian parameters, and impose the constraints on weak scale parameters by requiring that the Higgs mass is $125 \pm 1 \gev$ and that the VEV is $246 \gev$. This typically requires that dimensionful parameters are \order{T_C} and dimensionless parameters are \order{1} at the high scale. GW from high-energy PTs were considered in \refcite{Jinno:2015doa}.

A fraction of the latent heat from the PT could ultimately be released in collisions between bubbles, % SW: and sound waves from bubble nucleation, 
which result in striking GW signatures. This occurs at the bubble nucleation temperature, $T_N$, which is typically similar to the critical temperature, $T_N \lesssim T_C$, \ie the temperature at which the original ground state and emerging ground state are degenerate. We will calculate the nucleation temperature in order to calculate the peak frequency and the amplitude of the GW resulting from the singlet PT.

\section{Phase transitions in a temperature improved potential}\label{Sec:PT}

In this section we investigate whether the SM extended with a singlet can produce GW at a strongly first-order PT which could be detected by aLIGO. Acceptable low energy phenomenology, including standard Higgs properties and vacuum stability, is imposed. 
To achieve such a scenario we require the following cosmological history.
\begin{enumerate}
\item Higgs and singlet fields are in true, stable vacuum at origin at high temperature.

\item At $T\approx T_N \approx T_C \in (10^7,10^8) \gev$, the singlet acquires a VEV in a strongly first-order PT generating GW, potentially in reach of aLIGO. (The temperature was chosen to coincide with the peak frequency sensitivity in aLIGO (O5).)

\item At low temperature, the Higgs acquires a VEV, $\vev{H}\approx 246\gev$, resulting in the correct weak scale, Higgs mass, and satisfying constraints on Higgs-singlet mixing.
\end{enumerate}
We will calculate the critical and nucleation temperatures numerically as functions of the Lagrangian parameters. This is needed to calculate the frequency and amplitude of GW originating from bubble collisions. % SW: and sound waves. 
The first step is to include finite-temperature corrections to the effective potential. The one-loop finite-temperature corrections to the scalar potential have the form\cite{Patel:2011th,Quiros:1999jp}
\begin{equation}\label{Eq:VT}
\Delta V_T = \frac{T^4}{2 \pi^2} \left[
\sum_b J_B\left(\frac{m_b^2}{T^2}\right) + \sum_f J_F\left(\frac{m_f^2}{T^2}\right)
\right],
\end{equation}
where $J_B$ and $J_F$ are thermal bosonic and fermionic functions, respectively, and the sums are over field-dependent boson and fermion mass eigenvalues. We also add zero-temperature one-loop Coleman-Weinberg corrections\cite{Patel:2011th,Quiros:1999jp},
\begin{equation}
\Delta V_\text{CW} = \sum_i \frac{g_i m^2_i}{64 \pi^2} \left[\log \left(\frac{m^2_i}{\mu ^2}\right) - n_i \right],
\end{equation}
summed over massive particles, where $\mu$ is the renormalization scale, chosen to minimize large logarithms; $m_i$ is a field-dependent mass eigenvalue; $g_i$ is the numbers degrees of freedom associated with the massive particle; and $n_i=3/2$ for scalars and fermions and $5/6$ for massive gauge bosons (up to an overall sign for fermions). 

 Note that when one considers a PT in the singlet direction the only relevant masses are field dependent mass eigenvalues of both the CP even and CP odd scalar mass matrices as well as the charged Higgs. Also, there are no issues with gauge dependence. The final corrections to the finite-temperature effective potential are the Debye masses $\Delta V_D$ which result in the Lagrangian bare mass terms obtaining corrections of the form $\Delta  m_T^2 \propto T^2$\cite{Das:1997gg}. Thus, we consider the one-loop finite-temperature potential
\begin{equation}\label{Eq:Free_energy}
V = V_0 +\Delta V_D+ \Delta V_T + \Delta V_\text{CW}.
\end{equation}

\begin{figure}
    \centering
    \includegraphics[width=0.9\linewidth]{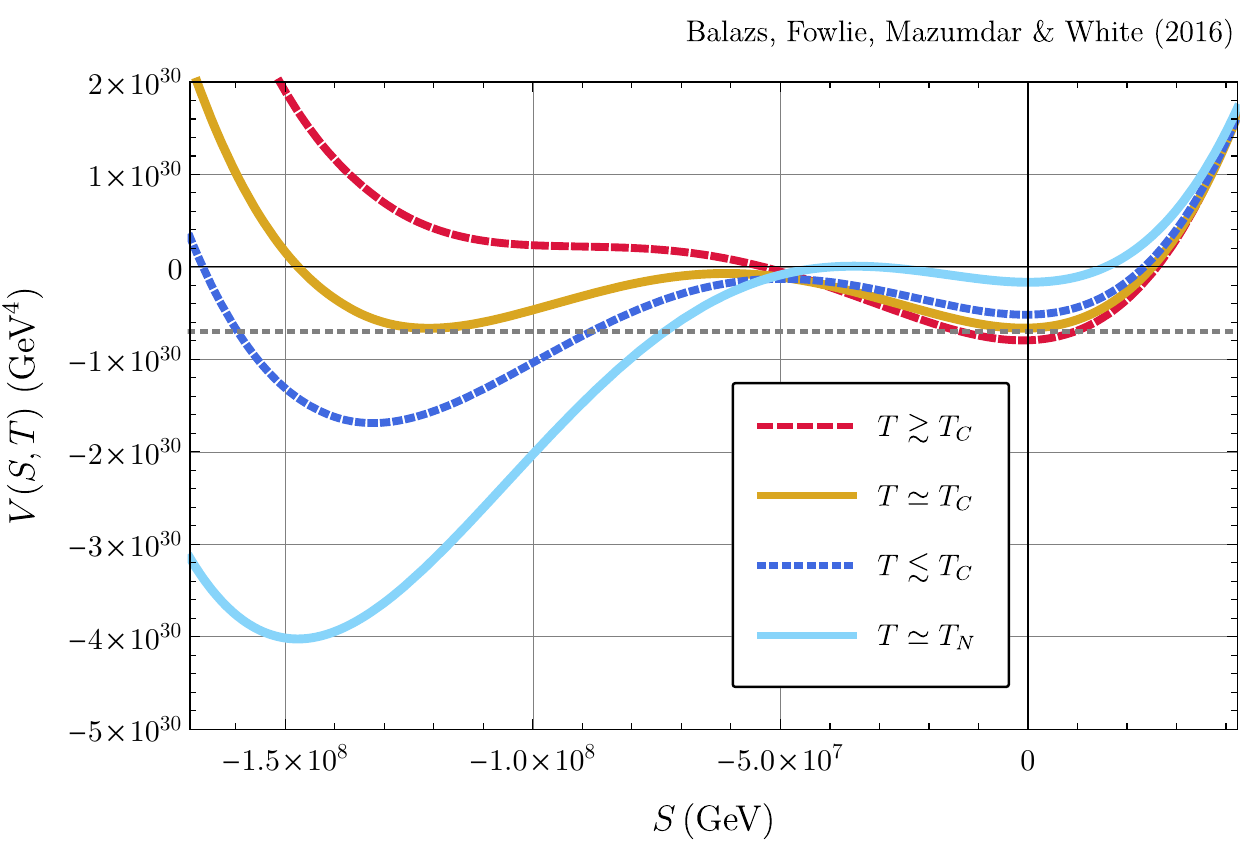} 
    \caption{The effective potential (\ie free energy) for benchmark SSM II, shown above, below and at the critical temperature, $T_C$, at which the minima are degenerate, and at the nucleation temperature, $T_N$.}
    \label{Fig:BM2}
\end{figure}

The conditions for a strongly first-order PT generating GW are that
\begin{enumerate}
\item There are at least two minima,
\begin{equation}\label{Eq:PT_minima}
\left.\pderiv{V}{S}\right|_\mathcal{F} = \left.\pderiv{V}{S}\right|_\mathcal{T} = 0.
\end{equation}
The calligraphic subscripts indicate the expression should be evaluated in the true ($\mathcal{T}$) and false ($\mathcal{F}$) vacua.\footnote{The vacua are degenerate at the critical temperature. We, however, always refer to the deepest minimum at zero temperature as the true minimum.} 
\item There exists a critical temperature, $T_C$, at which the two minima are
degenerate,
\begin{equation}\label{Eq:PT_degenerate}
V|_\mathcal{F} = V|_\mathcal{T}.
\end{equation}
This is illustrated in \reffig{Fig:BM2} for a benchmark point tabulated in \reftable{Tab:BM} by SSM I.
\item The order parameter at the critical temperature
\begin{equation}
\gamma \equiv \frac{\vev{S}}{T_C},
\end{equation} 
must be substantial (\ie \order{1}) in order to yield a strong first-order PT. The fact that $S$ is a gauge singlet means that we do not need to concern ourselves with subtleties involving gauge invariance\cite{Patel:2011th}.

\item Bubbles form, expand, % SW: resulting in sound waves, 
dominate the Universe and violently collide.

\end{enumerate}
For the first-order PT generating GW, we fix the critical temperature and order parameter, and solve for Lagrangian parameters at the high scale such that the conditions hold. 

The peak frequency and peak amplitude of the resulting GW are controlled by the nucleation temperature, $T_N$, which is the temperature at which a $1/e$ volume fraction (given by the Guth-Tye formula\cite{PhysRevLett.44.631}) of the Universe is in the true vacua. By dimensional analysis, this approximately occurs once 
\begin{equation}\label{Eq:p_time}
p(t) t^4 \approx 1,
\end{equation}
where $p(t)$ is the probability per unit time per unit volume that a critical bubble forms. As a function of temperature,
\begin{equation}\label{Eq:p_T}
p(T) \approx T^4 e^{\frac{-\se(T; S_\text{b}(r; T))}{T}},
\end{equation}
where $\se(T; S_\text{b}(r; T))$ is the Euclidean action evaluated along a so-called bounce solution. The Euclidean action is defined as
\begin{equation}
\se = 4 \pi \int_0 ^\infty r^2 \dint r \left[\left( \deriv{S(r)}{r} \right)^2 + V(S,T)\right], 
\end{equation}
and is a functional of the singlet field, $S(r)$. A bounce solution is a solution to the classical equation of motion for the singlet\cite{Coleman:1977py}. That is, we must solve 
\begin{align}\label{Eq:EoM}
\begin{split}
\frac{\partial ^2 S}{\partial r^2}+\frac{2}{r} \frac{\partial S}{\partial r} = \frac{\partial V(S, T)}{\partial S},\\
S^\prime(0) = 0, ~~~ S(\infty) = 0,
\end{split}
\end{align}
for $S_\text{b}(r; T)$, where the effective potential is defined in \refeq{Eq:Free_energy}. In a radiation dominated Universe, temperature and time are related by
\begin{equation}\label{Eq:T_time}
T^2 t = \sqrt{\frac{45}{16\pi^3}} \frac{\MP}{\sqrt{g_\star}},
\end{equation} 
where $g_\star \approx 100$ is the number of relativistic degrees of freedom and $\MP$ is the Planck mass. Combining \refeq{Eq:p_time}, \refeq{Eq:p_T}  and \refeq{Eq:T_time} results in the condition that the Euclidean action satisfies
% \footnote{This expression assumes that $V(S=0, T_N) = 0$. The form
% \begin{equation*}
% \frac{\se(T_N; S_\text{b}(r; T_N)) - \se(T_N; S = 0)}{T_N} \approx 
% 170 - 4\ln\left(\frac{T_N}{1\gev}\right) -2 \ln g_\star,
% \end{equation*}
% is manifestly invariant under $V \to V + \text{const.}$
% } 
\begin{equation}\label{Eq:TN}
\frac{\se(T_N; S_\text{b}(r; T_N))}{T_N} \approx
170 - 4\ln\left(\frac{T_N}{1\gev}\right) -2 \ln g_\star.
\end{equation} 
We solve for the nucleation temperature $T_N$ in \refeq{Eq:TN} by bisection, finding the bounce solution and the resulting Euclidean action for every trial temperature. To find a bounce solution, we approximate the bounce solution by perturbing about an approximate kink solution\cite{Akula:2016gpl}.

\section{Gravitational Waves}\label{Sec:GW}

The amplitude of GW from a first-order PT
depends on the wall velocity of a bubble, $v_w$; the latent heat released in the transition between the true and
false vacuum, $\Delta \rho$; the efficiency of the conversion of latent heat to GW; and the duration of
the transition. The latter is parameterized by
\begin{equation}\label{Eq:beta_exact}
\beta \equiv - \deriv{\mathcal{S}_4}{t} \bigg|_{t_N} = H_N \left[\deriv{\ln \se / T}{\ln T}\right] \frac{\se}{T} \bigg|_{T_N}
\end{equation}
where $\mathcal{S}_4 = \se / T$ is the four-dimensional Euclidean action for a bounce solution to
the equations of motion, $t_N$ is the nucleation time and $H = -\dot{T}/T$. 
% We used \refeq{Eq:T_time}. 
The characteristic timescale of the PT is $1/\beta$. We can approximate the timescale by\cite{Turner:1992tz,Hogan:1984hx}
\begin{equation}\label{Eq:beta}
\frac{\beta}{H_N} \approx \frac{\se(T_N)}{T_N},
\end{equation}
up to an $\order{1}$ factor. We solved the right-hand side in \refeq{Eq:TN}. We attempt to calculate $\beta$ by numerical differentiation of the action with respect to temperature in \refeq{Eq:beta_exact}; however, to reflect uncertainties in our calculation, we furthermore present results from varying the timescale of the PT between $1\leq \beta/H_N\leq 200$. The lower bound is from causality\cite{Giblin:2014gra} --- the characteristic size of a bubble cannot exceed a horizon --- and the upper bound is slightly greater than the approximation in \refeq{Eq:beta}.

% However, an exact time scale for the first order phase transition is rather difficult to compute 
% in general and one may have to resort to lattice simulations for a better estimate than the above estimation. In a thin wall approximation, 
% $\beta/H_N\sim 5/\epsilon$, where $\epsilon $ is the split in the vacuum energy density between true and false vacua. Typically, for 
% $0<\epsilon \leq 1$, $\beta/H_N\sim {\cal O}(100-1000)$. However, for a thick wall scenario, and in some extreme cases, one can obtain
% $\beta/H_N\sim {\cal O}(1)$\cite{Schwaller:2015tja}. Here we will not fix $\beta/H_N$, but we will allow it to vary from $1\leq \beta/H_N\leq 200$. 
% There are also other contributions to GW spectrum, such as plasma turbulences and bubble-wall 
% collisions~\cite{arXiv:1304.2433,Caprini:2007xq}, which we will not consider here.

The latent heat is parameterized by
\begin{equation}
\alpha \equiv \frac{\Delta \rho}{\rho_N} \quad\text{where}\quad \rho_N \equiv \frac{\pi^2 g_\star T_N^4}{30}.
\end{equation}
The denominator $\rho_N$ is the energy density of the false vacuum and $g_\star = 107.75$ is the number of relativistic degrees of freedom at the nucleation temperature $T_N$. 
The numerator, $\Delta \rho$,
is the latent heat in the transition between the true and false vacuum,
\begin{equation}
\Delta \rho = \left[V - \deriv{V}{T} T_N \right]_\mathcal{F}
             - \left[V - \deriv{V}{T} T_N \right]_\mathcal{T},
\end{equation}
evaluated at the nucleation temperature, where $V$ is the temperature improved scalar potential (\ie free energy) and subscripts indicate true ($\mathcal{T}$) and false ($\mathcal{F}$) vacua.

The bubble wall velocity --- a factor that influences the amplitude of GW ---
is slowed by friction terms arising from interactions with
particles in the plasma. In the high-scale PT that we are considering, because
there are fewer friction terms than in the EWPT in the SM, we expect that $v_w\approx 1$ in general.\footnote{In supersymmetric models, the
wall velocity of bubbles in an EWPT tends to be
heavily suppressed by strongly interacting scalars\cite{John:2000zq}. In the Standard
Model (SM), the wall velocity in an EWPT is significantly higher without these
friction terms. Thus, for a high-scale PT in the SSM, with even fewer friction terms, we expect $v_w \simeq 1$.}
The efficiency of converting latent heat into GW --- the final factor affecting GW --- is denoted by $\epsilon$. 
Because in our scenario $\gamma \gtrsim 1.75$ (\ie we consider a very strongly first-order PT), 
one finds that $\epsilon \approx 1$. We take $\epsilon=1$ throughout. 
% SW: for bubble collisions and sound waves.

Combining all the factors, from numerical simulations using
the so-called envelope approximation (see \eg \refcite{Jinno:2016vai} for an analytic calculation), the peak amplitude
of the GW strength, defined as the energy density per logarithmic frequency interval in units of the critical energy density of the Universe, due to bubble collisions measured today is given by
\begin{align}
\begin{split}
\Omega_\text{GW} \simeq 10^{-9} \cdot & \left(\frac{31.6 H_N}{\beta}\right)^2 \left(\frac{\alpha}{\alpha + 1}\right)^2\\
              {}& \epsilon^2 
              \left(\frac{4 v_w^3}{0.43 +v_w^2}\right)              
              \left(\frac{100}{g_\star}\right)^{\tfrac{1}{3}}.
\end{split} % \\
% \begin{split}
% \Omega_\text{SW} \simeq 10^{-9} \cdot & \left(\frac{5765 H_N}{\beta}\right) % \left(\frac{\alpha}{\alpha + 1}\right)^2\\
%              {}& \epsilon_\text{SW}^2 
%              v_w              
%              \left(\frac{100}{g_\star}\right)^{\tfrac{1}{3}}.
%\end{split}
\end{align}
% SW: from bubble collisions (BC) and sound waves (SW), respectively, 
where $g_\star=107.75$ in our model. The factors are \order{1} for a PT at a nucleation temperature $10^7\gev \lesssim T_N \lesssim 10^8\gev$. The peak amplitude is \order{10^{-9}} for $\alpha \simeq 1$ and $\gamma \simeq 2$. The aLIGO experiment, LIGO running phase O5, should be sensitive to amplitudes greater than about $\Omega_\text{GW} \gtrsim 5\times10^{-10}$ at about ${\cal O}(10)-{\cal O}(100)\,\text{Hz}$\cite{TheLIGOScientific:2016wyq,Thrane:2013oya}. 
% In our numerical calculations we assume that the wall velocity is equal to one. We typically expect the wall velocity to be supersonic and close to unity as there are no massive particle to provide friction in the plasma apart from the singlet (this is in contrast to say the MSSM during the EW PT where stops etc. provide a large source of friction dramatically dampening the wall velocity\cite{John:2000zq}). A precise calculation of the wall velocity we leave to future work. 

The peak amplitude observable today occurs at the peak frequency
\begin{equation}
f_0 \simeq 16.5 \,\text{Hz} \cdot \left(\frac{f_N}{H_N} \right) \left( \frac{T_N}{10^8\gev}\right) \left(\frac{g_\star}{100} \right)^{1/6}\\
% f_\text{SW} &\simeq 19.0 \,\text{Hz} \cdot \frac{1}{v_w}\left(\frac{\beta}{H_N} \right) \left( \frac{T_N}{10^8\gev}\right) \left(\frac{g_\star}{100} \right)^{1/6}\\
\end{equation}
% SW: for bubble collisions and sound waves, respectively, 
where $f_N$ is the peak frequency at the nucleation time,
\begin{equation}
f_N =\frac{0.62 \beta}{1.8-0.1v_w+v_w^2}.
\end{equation}
The peak frequency of GW from a PT coincides with aLIGO's maximum sensitivity at about $20\,\text{Hz}$ if the nucleation temperature is about $10^7\gev \lesssim T_N \lesssim 10^8\gev$\cite{Dev:2016feu}.  

\section{Vacuum stability}\label{Sec:Vacuum_Stabiliity}

After the discovery of the Higgs boson, and subsequent determinations of its mass, 
the stability of the SM vacuum was re-examined\cite{EliasMiro:2011aa,Alekhin:2012py,Degrassi:2012ry,Masina:2012tz,Xing:2011aa}. 
At large field values, the SM effective potential is approximately,
\begin{equation}
V_\text{eff}(h) = \frac12 \lambda(\mu \approx h) h^4,
\end{equation}
and for stability it is sufficient to insure that, given an initial value of the quartic coupling at low energy, the RG evolution is such that
the quartic coupling is positive at least until the Planck scale. 

The result is sensitive to low energy data --- notably the top-quark mass, Higgs mass and strong coupling --- in the quartic coupling's RGE. With present experimental data, however, it is believed that the quartic coupling turns negative at about $\MI \simeq 10^{10}\gev$, referred to as the SM Higgs instability scale. The SM Higgs potential is believed to be metastable; although we live in a false vacuum, the probability of tunneling to the true vacuum is negligible\cite{Degrassi:2012ry}.   

This instability can be remedied in simple extensions of the SM, including the SSM, which could alleviate it by modifying the beta-function for the quartic coupling (at one loop by a fish diagram) or by negative corrections to the Higgs mass. The latter imply that a Higgs mass of about $125\gev$, as required by experiments, could be achieved with a quartic coupling larger than that in the SM, and could be realised by tree-level mixing which should result in a negative correction, as eigenvalues are repelled by mixing\cite{EliasMiro:2012ay,Lebedev:2012zw}. A quartic coupling sufficiently greater than that in the SM could insure that the quartic coupling remains positive until the Planck scale, though it should remain perturbative until that scale. 

There are, however, additional stability conditions in the SSM, such as 
\begin{equation}\label{Eq:SSM_stability}
\lambda \ge 0, \quad \lambda_S \ge 0, \quad \text{and}\quad \kappa_2 \ge -2\sqrt{\lambda_S \lambda},
\end{equation}
that result from considering large field behaviour in the $H=0$, $S=0$ and $\lambda H^4 = \lambda_S S^4$ directions in field space. Note that if $\kappa_2$ is negative, the latter condition is equivalent to  $\lambda_\text{SM} \ge 0$, that is, the SM vacuum stability condition. In this case, stability cannot be improved by a threshold correction, though could be improved by modified RGEs (see \refapp{Sec:beta-functions}). Thus, we consider $\kappa_2 > 0$. To insure perturbative unitarity, we followed \refcite{Kang:2013zba}. Because in our solutions the Higgs and singlet are approximately decoupled, it resulted in a constraint that $\lambda_S \lesssim 4.2$ below the GUT scale.

% \subsection{Vacuum stability from Higgs-singlet mixing}

We insure that the mixing angle between the doublet and singlet is negligible, such that our model agrees with experimental measurements indicating that the Higgs is SM-like. There is, however, a residual threshold correction to the SM quartic. After eliminating the mass squared terms by tadpole conditions, the tree-level mass-squared matrix in the basis $(h, s)$ reads
\begin{equation}
M^2 = \begin{pmatrix}
\lambda v^2 & \kappa_1 + \kappa_2 v_S  \\
\kappa_1 v+\kappa_2 v_S v & (4 \lambda_S v_S + \kappa) v_S - \frac12 \frac{v}{v_S}\kappa_1 v \\
 \end{pmatrix}.
\end{equation}
The off-diagonal elements lead to mixing between mass and interaction eigenstates, described by a mixing angle
\begin{equation}
\tan \theta \approx - \frac{\kappa_1+ \kappa_2 v_S}{4 \lambda_S v_S + \kappa}\frac{v}{v_S} + \order{\frac{v^3}{v_S^3}}.
\end{equation}
As the mixing is small, we use the same notation for mass and interaction eigenstates. The mass eigenvalues are approximately
\begin{align}
m_h^2 & \approx \left(\lambda - \frac{(\kappa_1 + \kappa_2 v_S)^2}{v_S (4 \lambda_S v_S+\kappa)}\right) v^2,\\
m_S^2 & \approx v_S (4 \lambda_S v_S+\kappa) - \frac12 \frac{v^2}{v_S} \left(\kappa_1-\frac{2 (\kappa_1+\kappa_2 v_S)^2}{\kappa +4 \lambda_S v_S}\right),
\end{align}
neglecting terms $\order{{v^4}/{v_S^2}}$. As stressed in \refcite{EliasMiro:2012ay,Lebedev:2012zw}, in the limit $v/v_S \to 0$, the singlet only partially decouples. Whilst the mixing vanishes ($\tan\theta \to 0$), a negative tree-level contribution to the Higgs mass survives: 
\begin{equation}
m_h^2 = \left(\lambda - \frac{(\kappa_1 + \kappa_2 v_S)^2}{v_S (4 \lambda_S v_S+\kappa)}\right) v^2 \leq \lambda v^2 .
\end{equation}
Thus, the quartic coupling in the SM plus a singlet that achieves $m_h \approx 125 \gev$ is greater than that in the SM (or equivalently, there is a threshold correction to the quartic coupling in an effective theory in which the singlet is integrated out from the SM plus singlet), which improves the stability of the Higgs potential. That is,
\begin{equation}\label{Eq:quartic_boost}
\Delta \lambda = \frac{(\kappa_1 + \kappa_2 v_S)^2}{v_S (4 \lambda_S v_S+\kappa)} \geq 0 . 
%=  \frac{\kappa_1 + \kappa_2 v_S}{v} \cdot \left|\tan\theta\right| \approx \frac{m_S^2}{v^2}\tan^2\theta.
\end{equation}
If $\kappa \to 0$ and $\kappa_1 \to 0$, $\Delta \lambda \to \kappa_2^2 / 4 \lambda_S$, reproducing the expression in \refcite{EliasMiro:2012ay,Lebedev:2012zw}. Substantial $\kappa_1$ in the numerator or cancellations involving $\kappa$ in the denominator could, however, help generate a sizable threshold correction.

There are, however, subtleties: the conditions in \refeq{Eq:SSM_stability} were necessary, but insufficient for stability. For example, in \refcite{EliasMiro:2012ay} it was shown that for a \z{2} symmetric potential and renormalization scales $\mu \lesssim M_S$, if $\kappa_2 > 0$, the SM vacuum stability condition,
\begin{equation}\label{Eq:SM_stability}
\lambda_\text{SM} \equiv \lambda - \Delta \lambda  \ge 0
\end{equation}
is required to avoid deeper minima in the $S=0$ direction. We thus require $\mu \lesssim M_S \lesssim \MI$, that is, that the singlet scale is less than the SM instability scale. This insures that although there is an instability scale at which the SM vacuum stability condition is broken,
\begin{equation}
\lambda_\text{SM}(\mu=\MI \gtrsim M_S) < 0,
\end{equation}
the vacuum may in fact be stable, as we may violate the SM vacuum stability condition at scales $\mu\gtrsim M_S$. We trust that lessons from the \z{2} symmetric case are applicable to our general potential in \refeq{Eq:VTree}. Thus, in this text, we describe our model as stable if the couplings satisfy the large-field conditions on vacuum stability in \refeq{Eq:SSM_stability} and the SM vacuum stability condition in \refeq{Eq:SM_stability} for $\mu \lesssim M_S \lesssim \MI$. We leave a detailed analysis to a future work.

\section{Numerical results}\label{Sec:Numerics}

As well as generating GW potentially within reach of aLIGO and improving vacuum stability, our models must satisfy low-energy experimental constraints on the weak scale (\ie the $Z$-boson mass), the Higgs mass and Higgs-singlet mixing, and be free from Landau poles below the GUT scale. We fixed an order parameter, $1.75 \lesssim \gamma \lesssim 5$, and a critical temperature of $10^7\gev \lesssim T_C \lesssim 10^8\gev$.

We included low-energy constraints by building two-loop renormalization group equations (RGEs) in \code{SARAH-4.8.2}\cite{Staub:2013tta} by modifying the \code{SSM} model and constructing a tree-level spectrum generator by finding consistent solutions to the tree-level tadpole equations and diagonalizing the weak-scale mass matrix. Our spectrum-generator guaranteed the correct weak scale by tuning the Higgs mass parameter in the tree-level tadpole equations. To approximately satisfy limits on Higgs-singlet mixing from hadron colliders \see{CMS-PAS-HIG-12-045}, we required a tiny mixing angle between Higgs and singlet scalars, $\tan\theta \le 10^{-6}$. We tuned the Higgs mass by bisection in the Higgs quartic such that $m_h = 125 \pm 1\gev$. We found simultaneous solutions to the low-energy constraints and GW requirements by iterating between the weak scale and the critical temperature. 

\begin{figure}%[htbp!]
    \centering
    \includegraphics[width=0.99\linewidth]{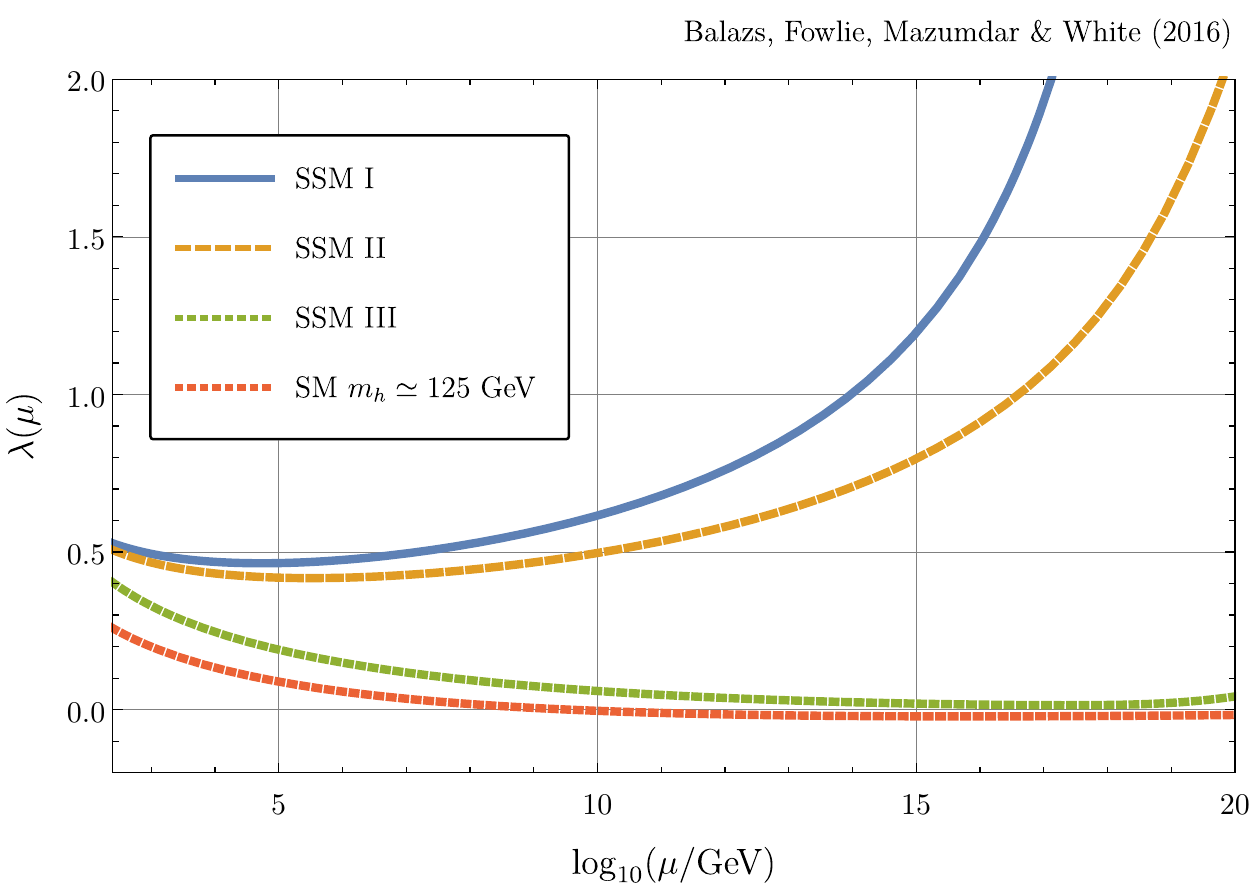} 
    \caption{Running of the Higgs quartic $\lambda$ in the SM and for our solutions in the SSM. All lines correspond to $m_h \simeq 125\gev$.}
    \label{Fig:lambda_Q}
\end{figure}

\begin{table*}

\begin{ruledtabular}
\begin{tabular}{lcccccccccccc}
Point & $M_S^2$/GeV${}^2$ & $\lambda_S$ & $\kappa$/GeV & $\kappa_1 $/GeV & $\kappa_2$ & 
$\lambda$ & $m_S$/GeV & $\gamma$ & $T_C$/GeV & $T_N / T_C$  & $\beta / H_N$ & $\Omega_\text{GW}$\\
\hline
SSM I & $4.2 \cdot 10^{14}$ & 0.064 & $2.1 \cdot 10^{7}$ & $-4.9 \cdot 10^{5}$ & 0.14 & 0.53 & $4.5 \cdot 10^{7}$ & 2.8 & $3.7 \cdot 10^{7}$ & $0.44$ & $118$ & $1.3\cdot 10^{-9}$\\
SSM II & $6.9 \cdot 10^{14}$ & 0.073 & $2.8 \cdot 10^{7}$ & $-7.3 \cdot 10^{5}$ & 0.15 & 0.51 & $5.5 \cdot 10^{7}$ & 2.9 & $4.2 \cdot 10^{7}$ & $0.45$ & $110$ & $1.3\cdot 10^{-9}$\\
SSM III & $1.3 \cdot 10^{15}$ & 0.13 & $7.4 \cdot 10^{7}$ & $-1.4 \cdot 10^{6}$ & 0.09 & 0.40 & $1.3 \cdot 10^{8}$ & 2.3 & $8.2 \cdot 10^{7}$ & $0.35$ & $45$ & $6 \cdot 10^{-9}$ \\
\end{tabular}
\end{ruledtabular}
\caption{\label{Tab:BM}Benchmark points, at the scale $Q=250\gev$, that exhibit GW potentially in reach of aLIGO (O5), vacuum stability, and acceptable low-energy phenomenology. The peak amplitudes were calculated for $\beta / H_N$ calculated numerically from in \refeq{Eq:beta_exact}.}
\end{table*}

\begin{figure}
    \centering
    \includegraphics[width=0.99\linewidth]{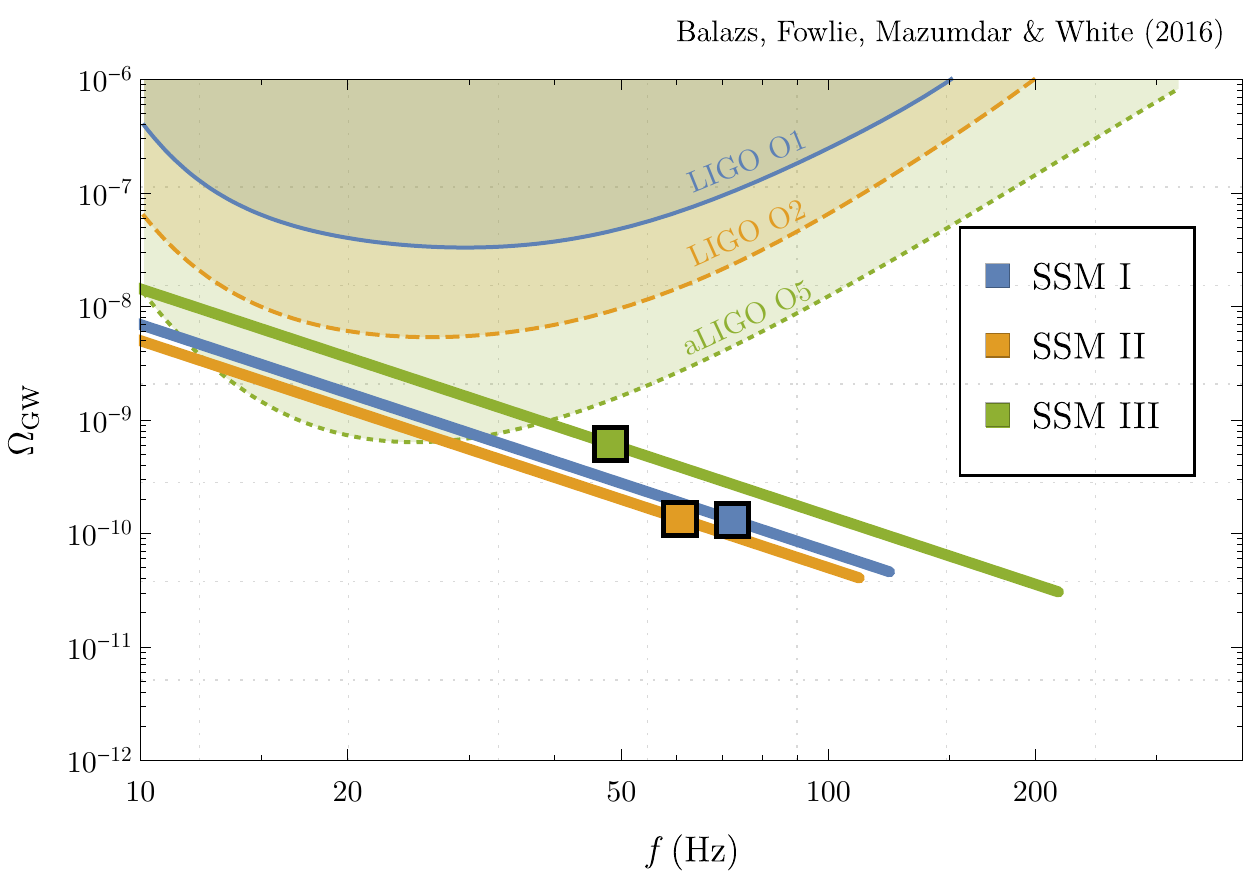} 
    \caption{Peak amplitudes and frequencies of GW for our SSM benchmark points from our approximate numerical calculation of $\beta / H_N$ (squares), with uncertainty represented by varying between $\beta / H_N = 1$ and $\beta / H_N = 200$ (lines). 
    The shaded regions indicate LIGO sensitivities during various phases of running\cite{TheLIGOScientific:2016wyq,Thrane:2013oya}.     
    All lines intersect the sensitivity of aLIGO (LIGO running phase O5). }
    \label{Fig:GW}
\end{figure}

In \reftable{Tab:BM} we present three benchmark points with GW amplitudes potentially within aLIGO (O5) reach, acceptable zero-temperature phenomenology and a substantial threshold correction to the tree-level Higgs quartic for improved vacuum stability. 
The running of the Higgs quartic for our three benchmarks and in the SM are shown in \reffig{Fig:lambda_Q}, demonstrating that for our benchmarks, the quartic coupling remains positive below the Planck scale, unlike in the SM.
Note that the running of the Higgs quartic coupling is sensitive to the precise values of the top Yukawa, $y_t$, and the strong coupling, $g_3$. The experimental measurements for $y_t$ and $g_3$ were boundary conditions at $Q\approx10^7\gev$; this introduced an error of up to about $3\%$ in their weak scale values for our benchmarks. As such the running for SSM III is pessimistic; its quartic running is probably steeper. For benchmark SSM I, the quartic coupling hits a Landau pole above the GUT scale.  We illustrate that our benchmark points result in peak amplitudes and frequencies of GW potentially within reach of aLIGO (O5) in \reffig{Fig:GW}. However, note that here we have varied $1\leq  \beta/H_N\leq 200$.

%{\bf Could we some how clarify the width we see in the plot, is it fiducial?. I believe it should be a line, am I correct ?, may be we show up to 100, if this looks better for selling our product.}

We selected our benchmarks from thousands of solutions found by Monte-Carlo (MC) sampling SSM parameters at the GW scale, $Q = T_C$, from the intervals
\begin{align}
\begin{split}
10^{-8}\gev \le & |\kappa_1| \le 10^{8}\gev\\
10^{-8} \le & \kappa_2 \le 2\\
10^{12}\gev^2 \le & M_S^2 \le 10^{18}\gev^2\\
10^{7}\gev \le & T_C \le 10^{8}\gev\\
2.3 \le & \gamma \le 3
\end{split}
\end{align}
We traded the Lagrangian parameters $\kappa$ and $\lambda_S$ for $T_C$ and $\gamma$ by solving \refeq{Eq:PT_minima} and \refeq{Eq:PT_degenerate}, and $\lambda$ and $\mu^2$ by requiring correct Higgs and $Z$-boson masses. A substantial fractional of our MC solutions could exhibit GW in reach of aLIGO; however, calculating the amplitude of GW accurately requires a thorough lattice simulation.

When selecting our benchmarks, however, we found that if $\gamma \gtrsim 3$, the rate of tunnelling is sometimes too slow for a PT to dominate the Universe with this being the case more often as $\gamma$ approaches $5$. That is, it is impossible to satisfy condition \refeq{Eq:TN} for any temperature. This is consistent with \refcite{Profumo:2014opa}, in which no solutions with $\gamma >5$ were found. Since we desire a completed PT, we discarded solutions with an order parameter $\gamma \gtrsim 5$. This may, in fact, be optimistic, as \refcite{Profumo:2014opa} indicates that completed PTs with $\gamma \approx 5$ are rare and as we require a lower value of $\se/T$ since the nucleation temperature is five orders of magnitude higher than the EW scale (see \refeq{Eq:TN}). On the other hand, if the order parameter $\gamma \lesssim 2.3$, the amplitude of GW may be too far below aLIGO (O5) sensitivity for all but the most optimistic estimate of the peak amplitude. There is therefore a ``Goldilocks region'' for the strength of the PT, $2.3 \lesssim \gamma \lesssim 3$, for which GW could be observed at aLIGO. Thus, to roughly select GW amplitudes in reach of aLIGO, we sampled from $2.3 \lesssim \gamma \lesssim 3$.

\begin{figure*}[htbp!]
    \centering
    \subfloat[Higgs quartic and dimensionless $S^2 H^2$ coupling.]{
        \includegraphics[width=0.4\linewidth]{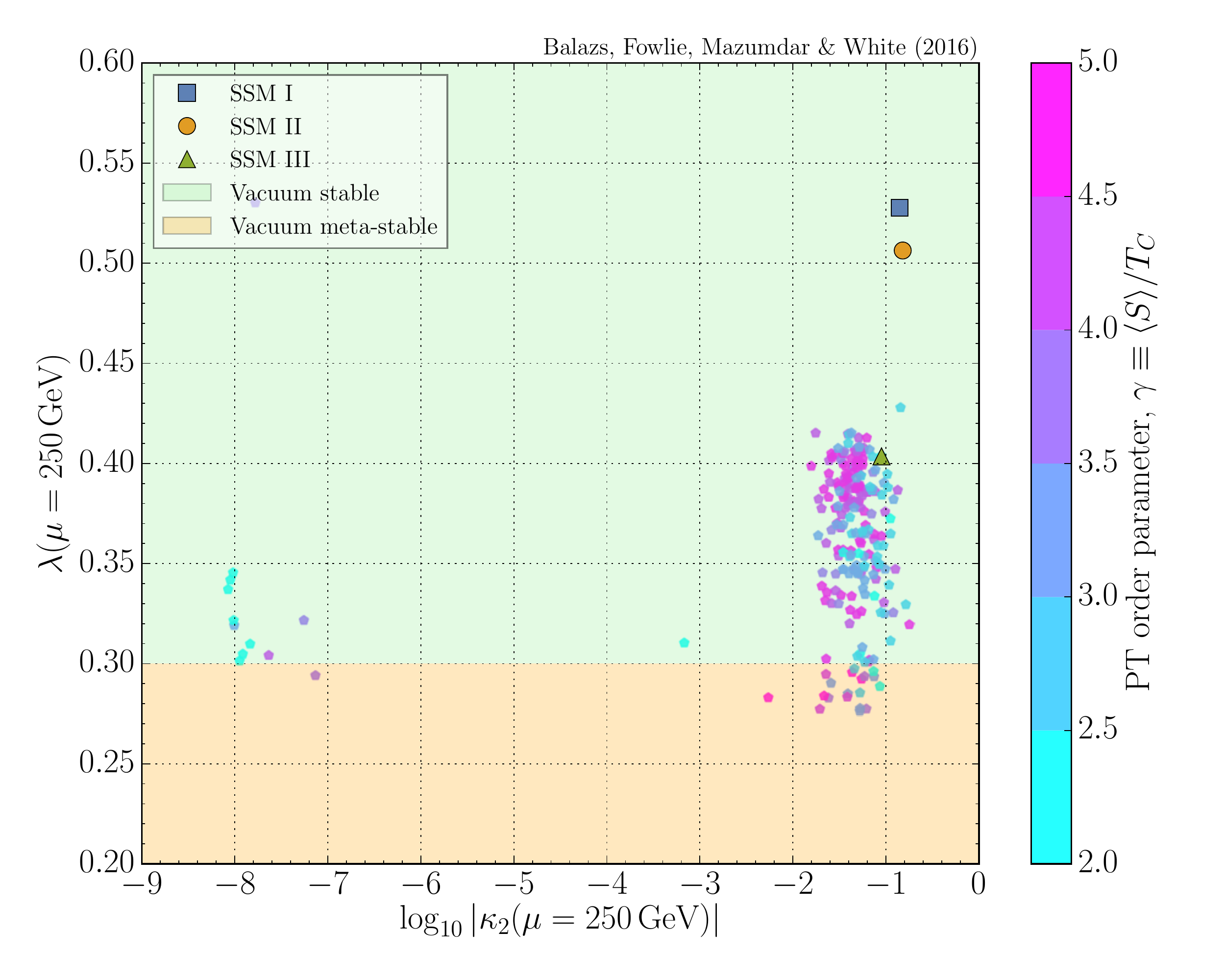}
        \label{Fig:lambda_k2}
    }    
    \subfloat[Higgs quartic and dimensionful $S H^2$ coupling.]{
        \includegraphics[width=0.4\linewidth]{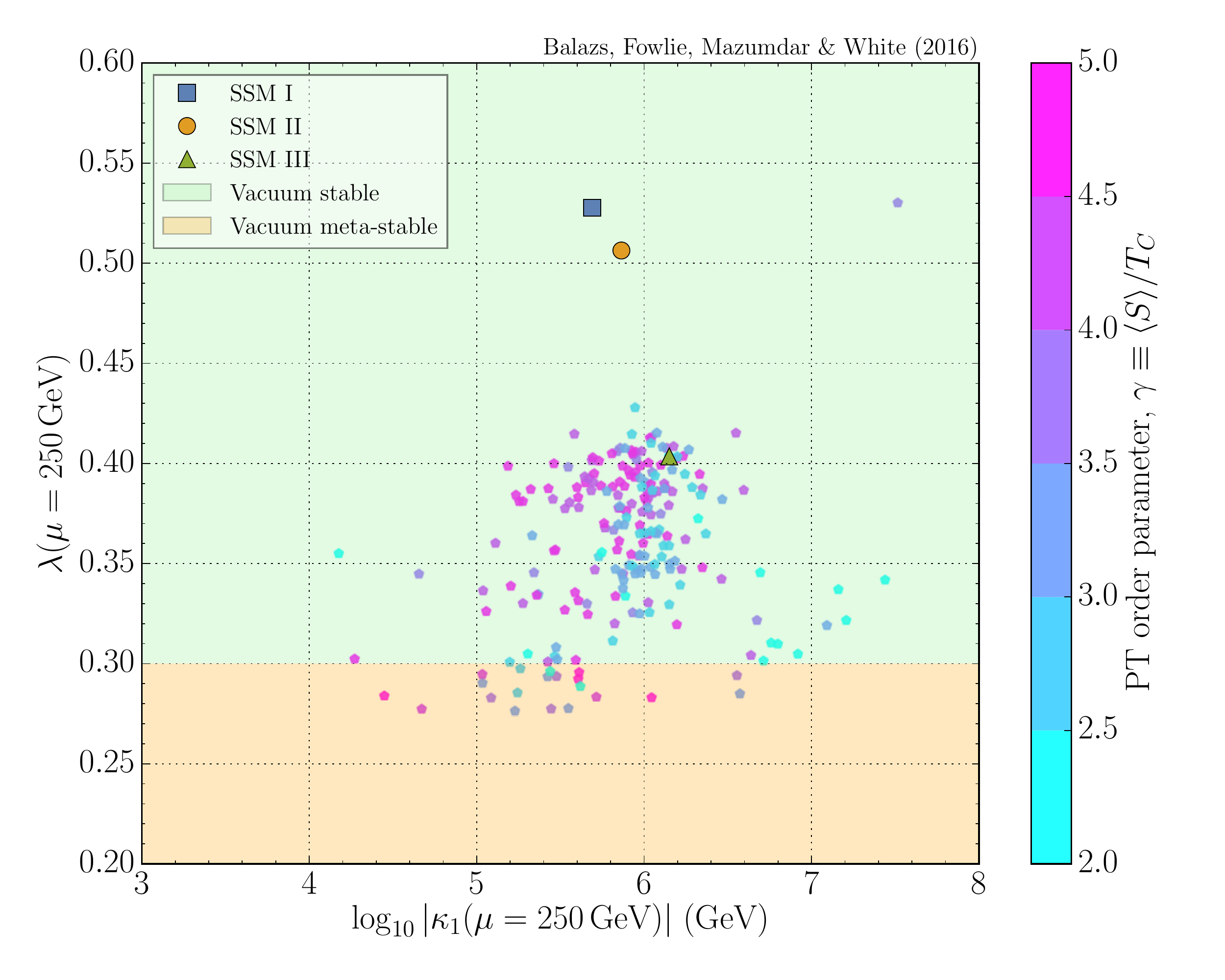} 
        \label{Fig:lambda_k1}
    }
        
    \subfloat[Higgs quartic and physical singlet mass.]{
        \includegraphics[width=0.4\linewidth]{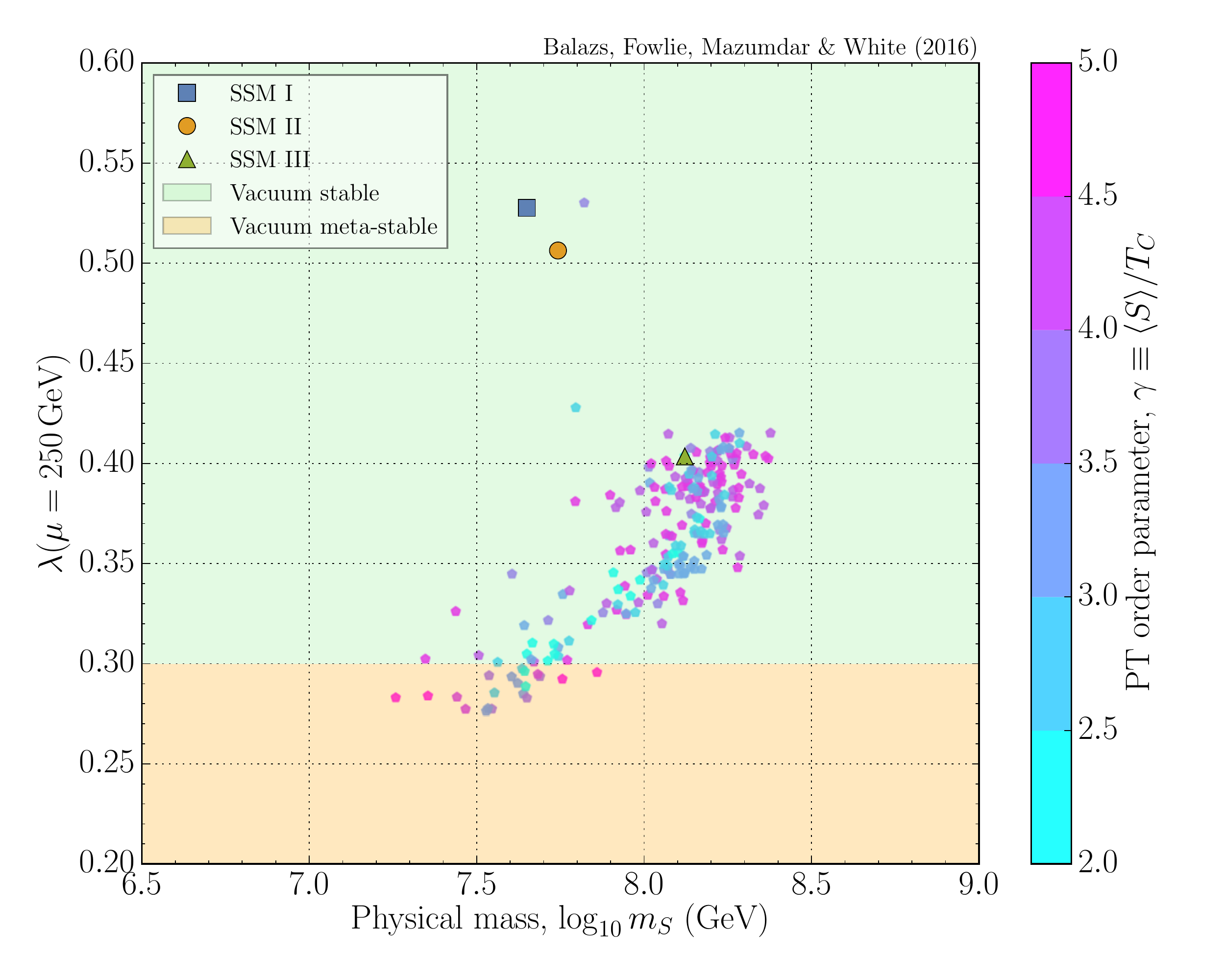}
        \label{Fig:lambda_ms}
    }
    \subfloat[The singlet-Higgs couplings]{
        \includegraphics[width=0.4\linewidth]{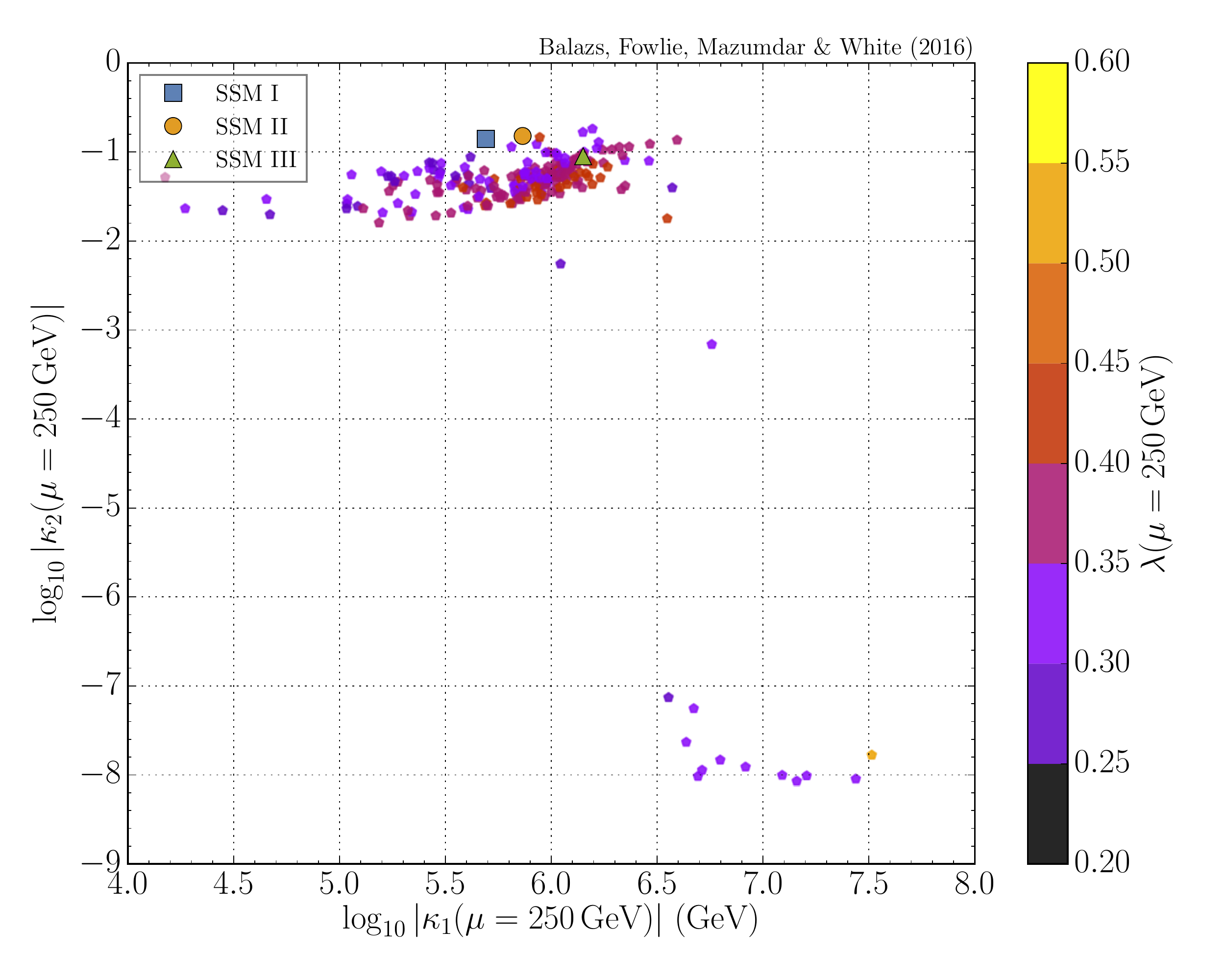} 
        \label{Fig:k1_k2}
    }
    \caption{Scatter plots of solutions in the SSM that exhibit strongly first-order PT at $T_C \in ( 10^7,10^8) \gev$,
    acceptable weak-scale phenomenology, and no Landau poles below the GUT scale. For the benchmark points shown, in addition, we checked that the PT results in GW signatures are potentially within reach of aLIGO (O5).}
    \label{Fig:scatter}
\end{figure*}

We scatter our MC solutions in \reffig{Fig:scatter}. We find that moderate Higgs quartics of $\lambda \sim 0.35$ are common, although there are outliers at $\lambda\gtrsim 0.4$. We see in \reffig{Fig:lambda_k2} that the dimensionless singlet-Higgs coupling is moderate, $\kappa_2 \lesssim 0.1$. We find, unsurprisingly, in \reffig{Fig:lambda_k1} and \reffig{Fig:lambda_ms} that dimensionful parameters are similar to the critical temperature, $m_S \sim \kappa_1 \sim T_C \sim 10^7\gev$. The Higgs-singlet couplings appear correlated in \reffig{Fig:k1_k2}. This is likely due to the fact that the Higgs-singlet mixing angle is reduced for $\kappa_1 \sim - 2 \kappa_2 v_s$. The sizes of the Higgs-singlet couplings are related to the threshold correction in \refeq{Eq:quartic_boost}, which we require to be moderate. There exist points with Higgs quartic larger than in benchmark SSM I that may suffer from Landau poles in the Higgs quartic below the GUT scale.

\section{Discussion and conclusions}\label{Sec:Summary}

GW detectors, such as LIGO, are a novel way of probing new physics.  In this work, we studied the detectability of primordial GW in the context of the SM augmented with a single real scalar field that is a singlet under all SM gauge groups.  The scale of the scalar singlet (its mass and VEV) was motivated by vacuum stability to be $10^7$-$10^8\gev $.  We have shown that, with this scale, the singlet dynamics leads to a strongly first-order PT that generates GW potentially within reach of aLIGO (LIGO run phase O5).  Selected from a wide sample over the parameter space, we presented three benchmark points with detailed calculations of the peak GW frequency and amplitude, demonstrating that for optimistic estimate of the peak frequency and amplitude, they lie within aLIGO sensitivity.  The most optimistic scenario, of course, arises for $\beta/H_N \sim {\cal O}(1)$.  

While it is known that eLISA is able to probe PTs at or near the EW scale, to our knowledge this work is the first to discuss a physical motivation for a PT to leave a relic background potentially detectable by aLIGO. Our result is due to the coincidence of aLIGO sensitivity with the EW instability scale.
% very strong  first-order 
% PT induced by a singlet extension of the SM, which also ensures that the EW vacuum is stable.
Indeed, the original analysis that proposed the existence of a heavy singlet leading to a tree-level boost in the Higgs quartic coupling, promoted the case where the mass of the singlet was $10^7$-$10^8\gev$\cite{EliasMiro:2012ay}. This is precisely in the region where the stochastic background is visible at aLIGO.  It should be stressed, though, that it is also possible to boost the stability of the vacuum with a lighter singlet.  

With planned LIGO running phases sensitive to GW amplitudes below $10^{-9}$, it is interesting to consider motivations for a PT at $10^7$-$10^8\gev$, which, on a logarithmic scale, lies about half way between the EW and the Grand Unification scales.  One exotic possibility is EW baryogenesis through a multi-step PT with the first transition at around $10^7$-$10^8\gev$ as proposed in \refcite{Inoue:2015pza}. This presents another intriguing possibility about physically motivated PTs occurring at such a high scale. This and other scenarios we leave to future work.

\begin{acknowledgments}
We thank Bhupal Dev, Eric Thrane and Peter Athron for helpful discussions. This work in part was supported by the ARC Centre of Excellence for Particle Physics at the
Terascale.  A.M. is supported by STFC grant ST/J000418/1.
% CB thanks his wife for her immense patience and understanding during the whole period of this project, and promises to take her out for a proper dinner immediately after the submission of the manuscript.  
\end{acknowledgments}

\appendix

\subsection{SSM $\beta$-functions}\label{Sec:beta-functions}

We generated beta-functions from our modified \code{SSM} model in \code{SARAH-4.8.2}\cite{Staub:2013tta}. The beta-functions for $\lambda_S$ and $\kappa_2$ were such that the quartics remained positive. The former is positive at one loop,
\begin{equation}
16\pi^2 \beta_{\lambda_S}^\text{1L} = \kappa_2^2 + 36 \lambda_S^2,
\end{equation}
though there are negative terms at two loop, and the latter is proportional to $\kappa_2$ at one loop,
\begin{equation}
\begin{split}
16\pi^2 \beta_{\kappa_2}^\text{1L} = \frac1{10} \kappa_2 (&-9 g_1^2- 45 g_2^2 + 60 \lambda + 60 y_t^2 \\
&+ 40 \kappa_2 + 120 \lambda_S ),
\end{split}
\end{equation}
and at two loop, thus at two loop it cannot change sign. There is, furthermore, an additional contribution to the beta-function of the SM quartic,
\begin{align}
\begin{split}
16\pi^2 \beta_\lambda^\text{1L} ={}& \frac{27}{100} g_1^4 + \frac9{10} g_1^2 g_2^2 + \frac94 g_2^4 - \frac95 g_1^2 \lambda\\
                                   & - 9 g_2^2 \lambda + 12 \lambda^2 +12 \lambda y_t^2 -12 y_t^4 
                                   + \kappa_2^2,
\end{split}
\end{align}
which could improve vacuum stability.

\bibliographystyle{h-physrev}
\bibliography{grav}
\end{document}